# Empirical effect of graph embeddings on fraud detection/ risk mitigation

Sida Zhou
Technology Centre, HENGCHANG
LITONG Investment
Management(Beijing) Co. Ltd.
Building 5, East District, Yard 10,
XIBEIWANG East Road, Beijing, China
sidazhou@gmail.com

*Abstract*—Graph embedding technics are studied with interest on public datasets, such as BlogCatalog, with the common practice of maximizing scoring on graph reconstruction, link prediction metrics etc. However, in the financial sector the important metrics are often more business related, for example fraud detection rates. With our privileged position of having large amount of real-world non-public P2P-lending social data, we aim to study empirically whether recent advances in graph embedding technics provide a useful signal for metrics more closely related to business interests, such as fraud detection rate.

I. INTRODUCTION

**There is a lot of money in P2P**

Peer-to-peer lending, abbreviated P2P lending, is where individuals can directly borrow money from another individual via online platforms. The online platform will act as the matchmaker between lenders and borrower, while simultaneously acting as risk mitigators by removing bad actors from the platform.

(Zhao2017) provides a comprehensive survey on the research about P2P lending. Most notably, in the past decade, there has been rapid development of online P2P lending platforms, examples of which include Prosper, LendingClub, and Kiva (Zhao2017). P2P lending could command $150 billion to $490 billion globally by 2020 (Stanley2015). Even with a fraud rate of 1%, the monetary value is still in the billions, hence any small improvement on fraud detection is worthwhile investigating.

**Purpose of fraud detection**

Making the loan, aka giving money away, is easy. The hard part is to get the loan back with interest months or years later, companies spend a lot of resources on collecting back the debt. It's not atypical for companies to have millions of clients, hence it's unpractical to manually check the information of each and every client, automated systems has to be built. The main purpose of these detection systems is to identify general trends of suspicious/fraudulent clients and/or transactions. Automated fraud detection is primarily happening on the borrower's side, since it's the borrower who pays interests, and is the source of income both for the lenders and for the platform.

P2P lending mostly happens on credit, there is no collateral, since the amount of loan is relatively small. Unlike traditional loans, the borrowers can potentially have non-existent or bad personal credit history. (Phua2013) surveys all published technical and review articles in automated fraud detection within 10 years' period until 2013, and (Phua2013) shows that the most common type of fraud is when credits are involved.

The goal is to identify intentional bad actors and downgrade or deny them the loan, in order to make better profits or to save money. Examples of bad borrowers: borrowers who tries to use forged or stolen identities; borrowers who take out loans with intention of disappearing with the money; borrowers who take out loans with intention to postpone repayments indefinitely.

**Importance of network analysis**

In modern society, vast majority of the population uses smart phones, and most social interactions have traces online. There are abundant social data that can provide valuable information on the task at hand. For example, a hypothesis is that people who makes repayments late tends to have friends who also make repayments late; or a person who have abnormal behaviors online might have his identity stolen etc.

One important way to utilize the social network data is via graph embedding methods. Graph embedding methods can automatically extract useful information of a network into an embedding space. We can then use this information as generated features together with a baseline fraud detection model to better accomplish our task at hand. Sections below will expand on the details.

We define the term "embedding space" loosely and we use it interchangeably with "feature space" and "vector space". We use the terms "graph" and "network" interchangeably. We use the terms "vertices" and "nodes" interchangeably.

Section II will explore related work in detail, and compare and contrast to our work. Section III will contrast our work to previous related work. Section IV will describe the methodology used in our work. Section V will discuss the results and conclude this paper.

## II. RELATED WORK

### A. Data for fraud detection research

**Problem with financial data**

(Phua2013) in 2013 conducted a "Comprehensive Survey of Data Mining-based Fraud Detection Research". (Phua2013) noted that there are often two main criticisms of data mining-based fraud detection research: 1) the dearth of publicly available real data to perform experiments on; 2) the lack of published well-researched methods and techniques. due to legal and competitive reasons (Phua2013). We agree on both points from our own experience.

**Problem with synthetic data**

To circumvent these data availability problems, one alternative is to create synthetic data. (Barse2003) uses a state machine to produce synthetic data. Synthesized data aim to preserves statistical distributions of certain features and properties, in order for the synthesized data to be representative. However, which features and properties are meaningful to preserve are unknown beforehand, and domain expertise or extensive training on real data is required (Lundin2002). For example, in synthesizing graph data, one might aim to preserve the degree distribution, while the research problem at hand is much more sensitive to PageRank of each node. In this case, the synthesize graph provides no value and will only mislead the research.

(Phua2013) noted that only 3 out of the 51 papers surveryed used simulated data but the data was either not realistic or the data and results were not explained (Phua2013). Hence, we conclude that synthetic data should be avoided if real data is available.

**Problem with public P2P data**

(Tarasenko) pointed out that some platform, namely LendingClub and Prosper have limited public data available through their websites and Kaggle (Tarasenko). There are also non-profit organizations who have some public data, e.g. theodi, Kiva, Zidisha (Tarasenko). However, these data are transactional in nature, for example the feature columns include "credit grade", "loan purpose", "home ownership", "DebtToIncomeRatio", " IncomeRange" etc. The data don't contain any information about the social relationships of the parties involved.

On the other hand, there are decent amount of public dataset on networks, with the purpose of perform research in social network and related fields. For example: SNAP's Large Network Dataset Collection from Stanford; KONECT (the Koblenz Network Collection) from University of Koblenz–Landau; The Index of Complex Networks (ICON) from University of Colorado Boulder. The social networks are especially relevant to the task at hand, for example the "Social circles from Facebook" or "Social circles from Google+", however these public data are often anonymized and not up to date and cannot be joined with the P2P lending data.

In our unique situation, we have the dataset that contains both the social information of the loan applicant and financial information and features of the applicant. Sections below will expand on the details.

### B. Related work on P2P lending

(Zhao2017) conducted a survey in 2017 on the research about P2P lending, section 2.1.2, secion 3.1.2 and section 3.2.1 of (Zhao2017) are most interesting to us. We have reviewed the most relevant studied to our work, and summarized the studies below.

(Freedman2008) (Lin2013) analyzed the public and private data from Prosper (prosper.com), a P2P lending platform, with a focus on social relationships. Prosper utilized social networking through groups and friends. Vaguely speaking, an individual may set up a group on Prosper and become a group leader. The group leader is to foster a community environment within the group so that the group members feel social pressure to pay the loan on time (Freedman2008). The relationships in question may be online or offline friends and/or acquaintances, or can be a part of social group, e.g. a Military Group or Alumni Group. An individual can be a member of only one group at a time (Lin2013).

To form a relationship on the platform, Prosper members with a verified email account can create a group and invite other Prosper members via direct invitation, or they can be invited to join existing groups via direct invitations (Lin2013). Incentives are set up for people to voluntarily form these relationships. For example, Prosper have distinction between levels of friends, Level 1 to Level 5, indicating how authentic this relationship is (Lin2013). This information is then displayed with the listing to strengthen the listings credibility.

(Freedman2008) tries to answer whether "social networks solve information problem". They engineered social network features manually using domain expertise, for example: "% In a Group", "% with Friends", "% w/ Group Leader Endorsement + Bid" etc. By using domain expertise, they also analyzed social groups, e.g. Military Group, Alumni Group etc.

(Freedman2008) stated that there is no doubt that having a social tie increases the funding rate, which is reasonable to us from our experience. However, the conclusion was with mixed signals, quoting (Freedman2008): "our data suggest that the market does not fully understand the signaling effect of social networks"

(Lin2013) investigates the statistical significance of various signals engineered using Prosper's social networks data. For example: "Total number of borrower's friends who are borrowers but not lenders", "Total number of borrower's real lender friends who bid on the borrower's listing" etc. They concluded that these social network based signals do have an effect.

(Etter2013) predicting the success of Kickstarter campaigns by using both direct information and social features extracted from Twitter. For the network component they preprocessed the data into a co-backers graph with project as nodes and backers as edges connecting the nodes. The features used are for example: "number of projects with co-backers", "proportion of these projects whose campaigns are successful" etc.

In studies summarized above, we note three things: 1) For Prosper, social relationships data acquired was formed on the platform was with specific intent to better one's profile, as oppose to data acquired naturally from Facebook or Twitter; 2) Most published studies only utilizes 1st degree relationships, as oppose to higher degrees; 3) Features were engineered by hand, requiring domain expertise, as oppose to automatically generating features.

### C. Graph embedding techniques

Graph embedding are a group of feature learning and dimension reduction technics in network analysis where nodes or edges from a graph are mapped to vectors of real numbers. Graph embedding are analogous to word embeddings from natural language processing. (PalashGoyal2017) (Hamilton2017) offer a great survey and review about the recent advances in graph embedding technics.

(PalashGoyal2017) categorizes embedding technics into: 1) Factorization based Methods, e.g. HOPE (Ou2016); 2) Random Walk based Methods, e.g. Node2vec (AdityaGrover2016), Deepwalk (Perozzi2014); 3) Deep Learning based Methods, e.g. SDNE (Wang2016). Sections below will elaborate on each of these methods.

Additional resources can be found on NLP Processing Lab at Tsinghua University (thunlpNRLPapers2018) (thunlpOpenNE2018), which offers a comprehensive and up-to-date "must-read papers" on network embeddings.

**Why graph embeddings now?**
There seem to be a growing interest in graph embedding in recent 3-4 year, as most papers were published in this period (thunlpNRLPapers2018) (PalashGoyal2017). Whereas most factorization based methods, involving matrices, were explored in the early 2000s (Yan2007). We think there are many potential reasons: new original ideas, e.g. Deepwalk (Perozzi2014); renewed interest in machine learning and deep learning; shift between analytical vs numerical paradigm; advancements in computing power and amount of available data. The real reason behind growing interest in graph embedding is an interesting topic to discuss.

### III. OUR WORK

(Phua2013) in 2013 notes that there are very few studies similar to our goal, stating that "virtually none uses spatial information" (Phua2013). Our goal is to examine empirically whether relational features generated by the graph embeddings provides a useful signal in fraud detection. Graph embedding provides information on the graph structure, which should be useful, however this information is noisy. Our work is to examine how the signal balances with the noise.

### IV. METHODOLOGY

Our pipeline consists of several different stages: 1) intrinsic feature engineering; 2) preprocessing graph data; 3) relational feature extraction using graph embedding; 4) feature addition; 5) classification. (See Figure 1)

Steps 1) and 5) are done routinely for classification tasks. Due to the advantage of our data, we can also extract graph features using steps 2) and 3) and concatenate all features in step 4) before doing the classification in step 5). (Tang2009) conducted somewhat similar experiment to ours, where they extracted latent social dimensions (similar to graph embeddings), then compared the predictive power of their classifier with and without the latent social dimensions, they saw an improvement when using latent social dimensions, see Figure 9 in (Tang2009).

### A. Intrinsic feature engineering

Our raw data is a combination of data from loan application form and data from user's mobile device that has been granted permissions to. All the data are anonymized. The data are for example information about spouse, emergency contact, referrers, employer, mobile device etc., and we refer to these as entities. Each entity has intrinsic and relational properties. Intrinsic properties are those, that are independent of the entity's relations, e.g. applicant's income, while the relational properties refer to other entities, e.g. applicant's relationship to other people.

For the intrinsic properties, standard feature engineering technics were applied, which consist of 1) standardization; 2) binarization; 3) binning; 4) one-hot encoding; 5) feature combination. After feature engineering, there are 558 intrinsic feature, which consist of approximately 1/3 features are based on telecommunications data, 1/3 are based on client's finance related data, and 1/3 are based on client's personal data. The data has ∼15% labelled as black (clients with bad behavior), and rest labelled as white (clients without bad behavior).

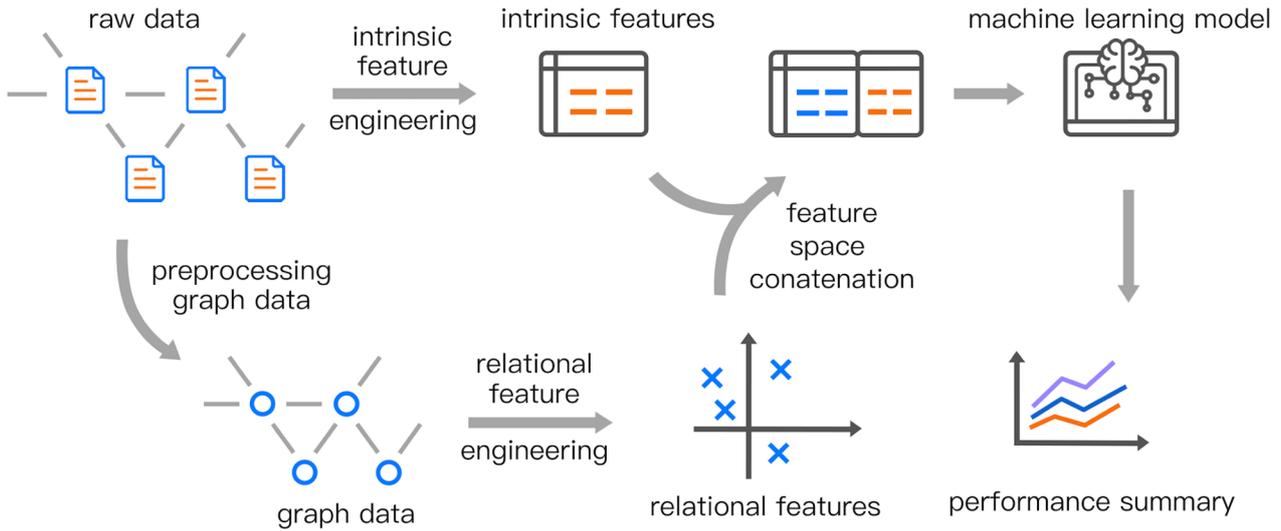

**Figure 1: Data pipeline from preprocessing of raw data to feature engineering and feature extraction to classification**

### B. Preprocessing graph data

As stated above, in each entity, there are various fields containing relationship information: e.g. emergency contacts, addresses, employer's information etc. These network data, together with network data from other sources are processed and stored in our graph database.

The total amount of data is large, therefore we subsample the data. We use top 10% of most suspicious neighborhoods within a selected province and randomly selecting a node within these neighborhoods as starting nodes, then we take the connected components of these nodes as our subgraph. The resulting subgraph is 3.8e6 nodes and 8.0e6 edges, cf. Table 1, this corresponds to 0.29e6 loan application forms processed. We also briefly considered the basic Forest Fire model (Leskovec2005), however Forest Fire model aimed to preserve properties less important to us, for example heavy-tailed in-out-degrees, whereas we considered the local neighborhood of each suspicious node important.

Our graph database contains multiple different entities, meaning different types of nodes, for example "staff", "client", "loan", "company" etc. The graph database also contains multiple different types of relationships, a.k.a. edges, for example "approve", "work_at", "married_to". Different types and nodes and edges together forms facts, not unlike knowledge graphs, e.g. "staff_X-approve-loan_Y", "client_A-work_at-company_A" and "client_A-married_to-person_B" etc. (see Figure 2)

Most literature on network analysis and graph embeddings assume that there are only one type of node and one type of edge (PalashGoyal2017). Literatures on knowledge graphs consider graphs with different types of nodes and edges, see for example section 3.4 in (Gardner2015). We choose to investigate graph embeddings. Methods summarized in (PalashGoyal2017) can only handle graphs with single type of nodes and edges, hence we need to preprocess our graph data into a graph with only one type of node and one type of edge.

There are multiple ways to preprocess our raw graph data, depending on our goals. Section 4.1 and Fig 3 in (Subelj2011) shows why graph preprocessing is important and demonstrate nicely how different graph data can look like, while representing the same event. As an example for clarification, the raw data (Figure 2a) can be be preprocessed into co-workers graph (Figure 2b) or any-relation graph (Figure 2c), the networks might or might not be useful depending on the circumstances. Note that Figure 2b and Figure 2c only have single type of node and edge, which is what we wanted.

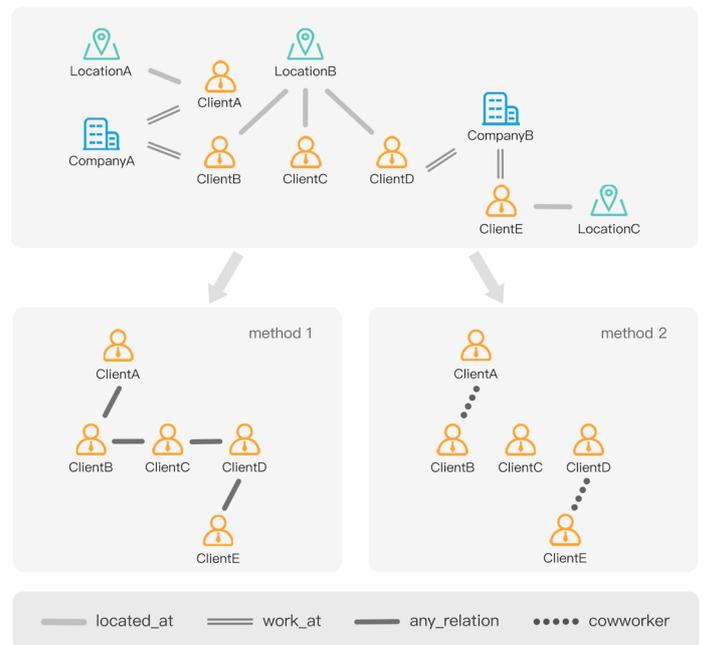

**Figure 2: representation of graph data in neo4j**

We choose each loan application to be a node in our graph. Edges are constructed combining all direct social relationships, namely 1) any contacts filled in the loan application form; 2) any contacts from user's mobile device that user granted permissions to. For our dataset, this strategy of constructing edges preserves majority of the edges and prunes entities that are irrelevant, first degree relationships are often more reliable and easier to handle. We ignore second degree relationships, for example adding an edge if two people work at the same company. Resulting graph is labelled hc-data in Table 1.

**Table 1: Dataset overview**

| Name | BlogCatalog | AstroPh * | HepTh ** | hc-data |
|---|---|---|---|---|
| $|V|$ | 10,312 | 18,772 | 9,877 | **2,236,218** |
| $|E|$ | 333,983 | 198,110 | 25,998 | **6,077,040** |
| Avg degree | 64.78 | 21.11 | 5.26 | **5.4** |
| Transitivity | 0.09 | 0.32 | 0.28 | **0.045** |
| Avg clustering | 0.46 | 0.63 | 0.47 | **0.035** |

\* discrepancy due to we treated as undirected graph, whereas (PalashGoyal2017) treated as directed graph
\*\* |V|, |E| didn't match with (PalashGoyal2017), likely due to different version of data
References for Table 1, please see (snap_node2vec_dataset2018) for BlogCatalog; (snap_dataset2018) for AstroPh and HepTh.

We see in Table 1, our data (hc-data) is on a different order of magnitude in term of its size. It is also more sparse, i.e. avg clustering is low. For definition and comparison between transitivity and avg clustering please see (Franceschet2018). (Leskovec2007) uses several datasets that have similarly low avg clustering, for example "EU email communication network" and "Patent citation network", they observed that most of these graphs densify over time. In other words, it is possible that our dataset is not sparse by nature, we just haven't collected enough data yet to densify the graph. On the other hand, caution is advised in measuring how similar graphs are, or how well one graph represent another graph, it is another field of study on its own (Leskovec2006).

### C. Relational feature extraction using graph embedding

A graph $G(V, E)$ is a collection of $V = \{v_1, ..., v_n\}$ vertices (a.k.a. nodes), where $n = |V|$ is the total number of vertices, and edges $E = \{e_{ij}\}$, where $i = 1 ... n$, $j = 1 ... n'$. A graph $G(V, E)$ can generally be represented by adjacency matrix $S$, with dimensions $n \times n$, with matrix elements $s_{ij} \in S$.

We are concerned with undirected and unweighted graphs. For unweighted graph, $s_{ij} = 1$ if an edge exist between vertex $i$ and vertex $j$, otherwise $s_{ij} = 0$. For undirected graphs, the adjacency matrix is symmetric, $s_{ij} = s_{ji} \forall i, j$.

A graph embedding is a mapping $f: v_i \to Y_i \forall i$, where $Y_i \in \mathbb{R}^d$, such that $d \ll |V|$, with $Y_i$ being the embedding vector of vertex $v_i$, with dimension $1 \times d$. The embedding matrix (a.k.a. relational features) for all vertices is $Y = \{Y_1; ...; Y_n\}$, with dimensions $|V| \times d$. The function $f$ preserves some proximity measure, defined on the graph, e.g. structural information (PalashGoyal2017). Note that collection of vertices $\{v_i\}$ or the adjacency matrix $S$ representation does not reside in a vector space, however the embedding representation $\{Y_i\}$ does reside. Therefore, for example, the distance measure, $distance(Y_i, Y_j)$ is defined and represents the proximity measure we define, whereas $distance(v_i, v_j)$ is not well defined and does not represent the proximity measure.

The research in graph embedding is to find this mapping function $f$. GEM paper (PalashGoyal2017) summarizes graph embedding methods extensively, and provides useful definitions, comparisons, complexity analysis, explores variable dependency and open source code (palash19922018) and more.

To extract the relational features, we used GEM (PalashGoyal2017) (palash19922018) as the primary source. We have selected few methods from GEM and studied in more detail: LLE, HOPE, node2vec and SDNE. We reused parts of GEM code (palash19922018) together with a hyperparameter optimizer. We reproduced, within the error margin, subset of the results in (PalashGoyal2017) for sanity check, namely Figure 2 and Figure 6 using BlogCatalog and AstroPh datasets, with most of the embedding methods.

We then investigated the behavior of the embedding methods with our data, which is substantially larger in size, $\sim 1GB$, see Table 1. The computing resource for the task was $40 cpus$, $400 GB\ memory$, no $GPU$. Using GEM implementation, we report following results:
- LLE did not converge
- HOPE threw Out-Of-Memory error
- node2vec successfully output embeddings
- SDNE threw errors on both tensorflow and theano backend, related to memory

On closer inspection, HOPE was implemented differently and less efficiently than the original paper (Ou2016), hence it consumed too much memory. In general, matrix based factorization/ eigendecomposition methods, including LLE(Fujiwara2017), are more memory intensive, and is less feasible to scale for larger datasets (AdityaGrover2016).

SDNE failure is probably due to lack of GPU, as the method is deep learning, autoencoder based. (Wang2016a) stated that it took one week to run for $0.1e6$ nodes (cf. our dataset has $2.2e6$ nodes), and complexity scales linearly with number of nodes.

Node2vec is the most feasible option out of methods investigated. Node2vec, as its name suggests, is analogous to word embedding word2vec (Mikolov2013): graph corresponds to document; node corresponds to word; edge corresponds to co-occurrence of words. There is a natural order of words in a document, but there isn't a natural order for nodes in a graph. Node2vec solves this problem by using biased random walks to generate the node sequences, and by balancing breadth-first search and depth-first search, node2vec preserves both the homophily and structural equivalence between nodes.

We used C++ version of node2vec by SNAP (snap_node2vec2018). We did randomized search over the hyperparameter space, to maximize the Mean Average Precision (MAP) of Graph Reconstruction, abbreviated as gr-MAP, see sections 5.2 and 6.1 in (PalashGoyal2017) for definitions. We chose Graph Reconstruction as the metric to maximize, because it represents how much information the embedding encodes of the original graph. We considered to maximize MAP of Node Classification, (section 6.4 in (PalashGoyal2017)) directly, as our goal was classification in the end. However, we decided Graph Reconstruction as the metric is more generalizable, since the output relational features can potentially be used for other tasks than classification.

The hyperparameters are (for definitions, see (AdityaGrover2016)):
- Embedding dimensions $d$
- Walks per node $r$
- Walk length $l$
- Context size $k$
- Return $p$
- In-out $q$

Randomized search was performed over the hyperparameter space. Evaluation criteria for each hyperparameter set is: we randomly sample 1024 nodes for evaluation, repeat for 5 times and calculate the MAP values for Graph Reconstruction. Our procedure is identical to section 6.1 in (PalashGoyal2017), with some hyperparameter differences. We then select multiple set of graph embedding with the highest MAP to be our multiple set of relational features, $X_{reln}$.

### D. *feature space concatenation*

Intrinsic features are engineered with domain expertise, and relational features are generated with graph embeddings, as some features are easier crafted by hand and others easier by using algorithms. We can combine intrinsic and relational features by using relational features as additional feature columns and let our classifier decide which set of features are important.

Attempts have been seen previously: (Yu2015) combines engineered features (intrinsic) and word embeddings (relational) with an outer product; (Tang2009) showed that combination of network (relational) and actors' (intrinsic) features is expected to lead to more accurate classification performance. (Damoulas2009) discusses this topic in more depth.

We have $n = |V|$ loan applications, for each sample we have m intrinsic features, such that intrinsic feature matrix $X_{intr}$ has shape n×m. Graph embedding generates $1 \times d$ embedding vector for each sample, hence the relational feature matrix $X_{reln}$ has shape $n \times d$. We can concatenate these feature matrices $X = [\ X_{intr}\ X_{reln}\ ]$, which has shape $n \times (m + d)$. For our data, $n = 26392, m = 558, d = 128 \ldots 256$.

### E. *Classification*

For this classification step, we outsourced the task to auto-sklearn (Feurer2015), which automatically choose a good algorithm and feature preprocessing steps and also set their respective hyperparameters. We had treated auto-sklearn as a black box.

We would use $X_{intr}$ on its own as the baseline performance of the classifier. Then we would use the concatenated feature set $X = [\ X_{intr}\ X_{reln}\ ]$ on the same classifier, in order to see the effects of adding $X_{reln}$. The hope is using feature set $X = [\ X_{intr}\ X_{reln}\ ]$ would yield better performance than using feature set $X_{intr}$.

For the performance metric, we used weighted f1-score. We varied the train-test ratio from 10% to 90%, see Figure 5a), expectation was higher train-test ratio would yield better performance. Hyperparameters were tuned such that weighted f1-score in training data were about the same as weighted f1-score in test data.

### V. RESULT AND DISCUSSION

There are many moving parts in our pipeline, we mainly focus on steps 3) relational feature extraction using graph embedding of the pipeline (see Figure 1).

There are reasons why we put less focus on other steps of the pipeline. For step 1) intrinsic feature engineering, we treat the intrinsic features as a baseline, we are only interested comparing the result with and without the relational features, hence the quality of the baseline and method used to obtain the baseline is irrelevant to our thesis. Nevertheless, the intrinsic features were engineered with great care in order to reflect the real world scenario. For step 2) preprocessing graph data, the result of our first efforts were discouraging enough that we didn't investigate further what would happen if preprocessing graph data would be done differently. For step 4) feature space concatenation, we chose to have the simplest case of combining feature spaces by concatenation, in order to have less moving parts. For step 5) classification, we chose to use auto-sklearn as a black-box that has little moving parts, with the expectation that if there are any signals strong enough, then auto-sklearn will be able to detect it.

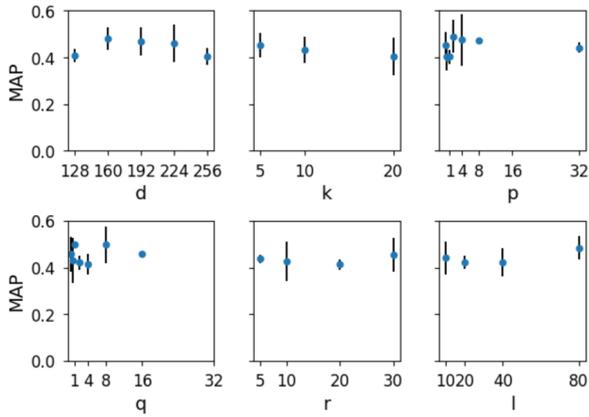

Figure 3: partial dependency of Graph Reconstruction MAP on node2vec hyperparameters of top 15% trials

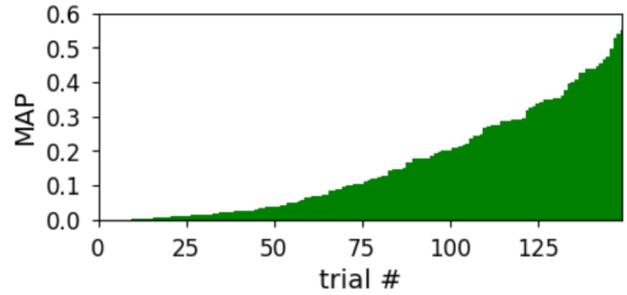

Figure 4: sorted MAP values of all trials

For step 3) Relational feature extraction using graph embedding, in total 150 hyperparameter sets were evaluated using node2vec, for search space see Figure 3. Totaling to 711 hours of computation, average 4.7h/trial. The time consumption is heavily dependent on the values of $d, r, l$. Larger values of of $d, r, l$ was not attempted due to time constrains. We sort 195 trials by MAP and plot top 15% trials that has largest MAP in Figure 3. Due to randomly combining hyperparameters, 50% of trials has MAP < 0.1, see Figure 4. We argue that certain random combination of the hyperparameters will yield suboptimal results and is meaningless and should be discarded.

Our data is much larger and sparser compared to (PalashGoyal2017), hence we decided to try a larger range of hyperparameters, hoping to see some unexpected trends. We expected to see a clearer trend in and walk length $l$ and embedding dimensions $d$, because large graph should require large $d$ to encode all necessary information, however this was not the case (see Figure 3). It is also unfortunate that no other interesting trends were observed. Finally, the few hyperparameter sets with the highest MAP is considered to be optimized and passed onto next steps.

It is fair to ask why steps 3) relational feature extraction and 5) classification are optimized separately, and not in a single step. The main purpose is to save computation time, we hypothesize that relational features with low gr-MAP will not improve the f1-score when added to the classifier, as it does not contain graph structure information, and will be just added noise. The search space is big, so we use gr-MAP as an indicator that indicates higher probability to get a better f1-score in classification step.

Figure 5a) shows the experimental results of auto-sklearn (Feurer2015) classifier using intrinsic features only vs using intrinsic+relational features together. The expectation is the relational features would contain graph structure information and hence will improve the predictive power of the classifier. The experiments showed that relational features did not improve the predictive power of the classifier.

One problem we encountered was attributed to overfitting in classification step, the f1-score was consistently higher in the training set than in the test test. We solved this problem by lowering the resources and training time available to auto-sklearn, initially from $time\_left\_for\_this\_task = 5h$, $per\_run\_time\_limit = 30min, ml\_memory\_limit = 30GB$, finally to $time\_left\_for\_this\_task = 6min$, $per\_run\_time\_limit = 8s, ml\_memory\_limit = 9GB$. With this limited resources, training set and test set error converged. Oversampling and undersampling were also attempted, to solve the unbalanced data issue, there was no improvements in the results.

When concatenated, the information of whether feature came from the relational feature space or the intrinsic feature space is lost. We have not found discussions on this information loss from published literature. We attempted, without theoretical justification, to do PCA on the relational feature space to mitigate this information loss. Relational feature space dimension of $d = 128 \ldots 256$ to PCA dimension of 100,50,10, only then we do the feature concatenation, there was no improvements in the results.

Classification results using intrinsic feature only vs intrinsic+relational features are presented in Figure 5a). However, additional experiments were also made to compare classification results using: relational features only; intrinsic +white noise. To summarize (see Table 2), weighted f1-scores of intrinsic+noise vs intrinsic+relational were similar. Intrinsic vs intrinsic+noise had similar f1-score. All additional experiments are further evidence of the relational feature did not carry useful information. We wanted to calculate how similar the relational features were to white noise, however a trivial method of calculating similarity of two multivariate distributions was not found.

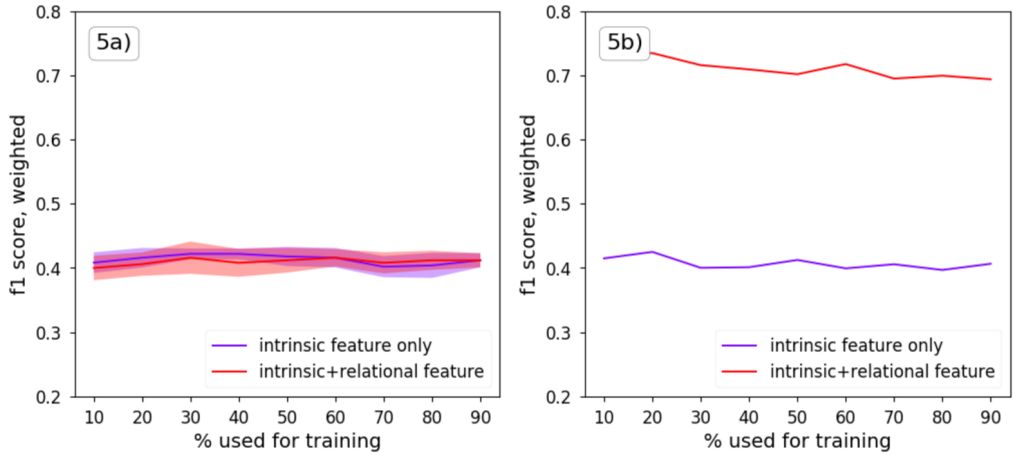

**Figure 5a):** Experimental results of performance of classifier with and without relational features; The error bars are due to different parameter sets used for auto-sklearn and/or different sets of relational features used
**Figure 5b):** Sanity checking classifier behavior when introduced strong but *fake* signals

**Table 2 – Additional experiment results, averaged over all experiments**

| Features used | Weighted f1-score |
|---|---|
| Intrinsic only | 0.41 |
| Intrinsic+relational | 0.41 |
| Intrinsic+noise | 0.42 |
| Relational only | 0.29 |

Figure 5b) started as an accident, a relational feature calculated from the label was used in predictions, hence it had high correlation with label, and hence gave the good result we desired to observe. This error was noticed PCA analysis, and the abnormal weight of this feature. We nevertheless present Figure 5b), because this error gives us confidence that if there are signals in the relational feature, then auto-sklearn will find it.

For the classification metric, (Lessmann2015) states that the AUC is not affected by class imbalance (Fawcett, 2006) As such, they are robust toward class skew in general. (Davis2006) states that a curve dominates in ROC space if and only if it dominates in PR space, and argues that PR curves are preferable when presented with highly-skewed datasets. The AUC of both ROC and PR were examined to be used metric for classification, instead of weighted f1. There were no interesting results.

## VI. CONCLUSION

Our goal is to examine empirically whether relational features generated by the graph embeddings provides a useful signal in fraud detection. We conclude that relational features generated by the graph embeddings does not provide a useful signal *under our experiment settings*. Nevertheless, we still wish to present our findings in this paper for future references.

We will emphasize that out conclusion is *under our experimental settings*, which are described in detail in this paper. We have successfully reproduced parts of the work of (PalashGoyal2017), which ensured that graph embedding technics were correct, however there are many other moving parts in our pipeline, which prevents us from drawing a more general conclusion. Perhaps the data itself does not contain the graph structural information we wanted to see, but to see how the algorithm behave with specifically our data is part of the experiment.

We have also outline many possible future directions that can be potentially fruitful to investigate in the section below.

## VII. FUTURE WORK

There are many things that can still be investigated in hope of finding better/any results, for example: Using directed and weighted graph data (as oppose to undirected and unweighted); Using different data, from different time period or different location; Using different hyperparameters in embedding and classification step; Using e.g. xgboost (instead of auto-sklearn) to have better control over parameter tuning; Using temporal information of the graph data; Examining the relationship between gr-MAP and weighted f1-score; search literature for experiments that have data more similar to ours.

On hyperparameter tuning, note that (Olson2017) empirically assessed 13 supervised classification algorithms on a set of 165 supervised classification datasets, and found that by tuning the hyperparameters there are about 5%~10% increase in accuracy at best (Fig. 3. in (Olson2017)). Hence our expectations should be set accordingly.

One interesting direction is to add fake signals to the graph, and see how much fake signal is needed before the graph embedding and the consequent classification manages to detect the signal. For example, we can attach triangles of black-black-black onto random nodes, i.e. three clients with bad behavior who all know each other, and see how many such groups need to be added for the algorithm to detect and classify correctly. This may also help in tuning the hyperparameters in graph embedding step, and aid in case studies, as we have control over the fake signals generated.

As mentioned in previous sections, when concatenated, the information of whether feature came from the relational feature space or the intrinsic feature space is lost. (Damoulas2009) states that concatenating all the features into a single feature space does not guarantee an optimum performance and it exacerbates the "curse of dimensionality" problem. (Damoulas2009) also states that the problem of classification in the case of multiple feature spaces or sources has been mostly dealt with in the past by ensemble learning methods, namely combinations of individual classifiers. So, another interesting direction is to investigate ensemble methods based on reference (Damoulas2009).

## IX. APPENDIX

A note for interest, for typical values of AUC, check out (Lin2015) (Sundarkumar2015) values (Nanni2009)

A note for interest, Table 4 from (AdityaGrover2016) and Table 1 from (Wang2017) are performance results of various graph embedding methods. While the numbers not directly comparable, the ordering seems to be consistent between the references.

A note on SDNE: Apart from scalability issue, we fail to see the justification of using the middle layer as the graph embedding $Y = Y^{(K)}$, referring to Alg. 1 in (Wang2016). SDNE is unlike word2vec (Mikolov2013) where the network is shallow and algorithm which produce word embeddings have better theoretical foundations. In SDNE encoder and decoder and the learned weights carries significant information in addition to the middle layer, hence we conclude using the middle layer $Y^{(K)}$, referring to Alg. 1 in (Wang2016) alone as graph embedding does not represent the graph well.

A note for interest, nearly every network we have examined, there is a substantial fraction of nodes that are barely connected to the main part of the network. For example, the Epinions network has 75,877 nodes and 405,739 edges, and the core of the network has only 36,111 (47%) nodes and 365,253 (90%) edges. (Leskovec2008)